\documentclass[aps,prc,superscriptaddress,twocolumn]{revtex4}

\usepackage{amssymb}
\usepackage{graphicx}
\usepackage{epsfig}%
\usepackage{bm}
\usepackage{hyperref}
\usepackage{mathrsfs}
\usepackage{multirow}
\usepackage{footmisc}
\usepackage{threeparttable}
\usepackage{booktabs}

\begin{document}

%
%
%
%
%

\title{In-medium $\mbox{NN}\to \mbox{N}\Delta$ cross section and its dependence on effective Lagrange parameters in isospin-asymmetric nuclear matter}

\author{Ying Cui}
\email{yingcuid@163.com}
\affiliation{China Institute of Atomic Energy, Beijing 102413, China}

\author{Yingxun Zhang}
\email{zhyx@ciae.ac.cn}
\affiliation{China Institute of Atomic Energy, Beijing 102413, China}
\affiliation{Guangxi Key Laboratory Breeding Base of Nuclear Physics and Technology, Guangxi Normal University, Guilin 541004, China}

\author{Zhuxia Li}
\affiliation{China Institute of Atomic Energy, Beijing 102413, China}

\begin{abstract}
The in-medium $NN\rightarrow N\Delta$ cross section and its differential cross section in isospin asymmetric nuclear medium are investigated in the framework of the one-boson exchange model by including the isovector mesons, i.e., $\delta$ and $\rho$ mesons. Our results show that the in-medium $NN\rightarrow N\Delta$ cross sections are suppressed with density increasing, and the differential cross sections become isotropic with the density increasing at the beam energy around the $\Delta$ threshold energy. The isospin splitting on the medium correction factor, $R=\sigma_{ NN\rightarrow N\Delta}^*/\sigma_{NN\rightarrow N\Delta}^{\text{free}}$ is observed for different channels of $NN\to N\Delta$, especially around the threshold energy for all the effective Lagrangian parameters. By analyzing the selected effective Lagrangian parameters, our results show that the larger  effective mass is, the weaker medium correction $R$ is.
\end{abstract}

\date{\today}


\pacs{Valid PACS appear here}

%

%

\maketitle

\section{Introduction}
The isospin dependence of in-medium $NN$ cross sections is a subject of much interest in the field of intermediate energy neutron-rich heavy ion collisions (HIC), since it can influence the predictions of reaction dynamics, collective flow, stopping power, and particle productions in the simulation of heavy ion collisions\cite{Chen1998, JYLiu01, QingfengLi2001, zhang07,lehaut10,lopez14,lwchen00,YJWang16}. By comparing the HIC  experimental data to the transport model calculations, the information of in-medium $NN$ cross section and equation of state (EOS) can be indirectly extracted. Principally, both the mean field (or EOS) and nucleon-nucleon cross section in the transport models should be determined by the same effective Lagrangian or effective interaction. However, the mean field potential and nucleon-nucleon cross section in varieties of transport models are treated independently due to the complexity of transport equations, and in particular their dimensionality. Especially, the solution of collision integral is not sought directly, but rather through Monte-Carlo cascade method in which the in-medium nucleon-nucleon scattering cross sections are adopted and their correction factor is ad hoc determined by fitting the related heavy ion collision observables. Thus, the direction of further improving the transport models in theory is to consider the mean field and nucleon-nucleon cross section consistently, it naturally requires to understand the relation between the in-medium $NN$ cross section and the EOS (or the nuclear matter parameters).

There are lots of efforts have been made to investigate the in-medium $NN$ elastic cross section and its isospin dependence by using the microscopic approaches \cite{Liguoqiang1993,Caiyanhuang1996,QingfengLi2000,Zhanghongfei2010}. In transport models, the isospin dependent medium correction factor $R=\frac{\sigma^*}{\sigma^{\text{free}}}=(\frac{m^*}{m})^2$ for the elastic $NN$ cross section has been adopted in the isospin dependent Boltzmann Uhling-Uhlenbeck (IBUU) and Lanzhou quantum molecular dynamics (LQMD)\cite{Persram2002,BALi05,Feng2012} models, and  phenomenological forms also have been applied in the different version of quantum molecular dynamics model (ImQMD, UrQMD), such as $R=(1-\alpha \rho/\rho_0)$ \cite{Zhangyingxun2007}, $R=F(\rho,p)$ \cite{Lipengcheng2018,YJWang16}, and $\sigma^*=\sigma_0\tanh(\sigma^{free}/\sigma_0)$ in Boltzmann Uhling-Uhlenbeck models (pBUU)\cite{Danielewicz2000}. However, there were few theoretical works to discuss the relation between the in-medium $NN\rightarrow N\Delta$ cross section and the EOS parameters, which becomes more and more important for further developing the transport models to study physics around the $\Delta$ threshold energy. Especially, with the urgent requirements on constraints of symmetry energy at suprsaturation density.

Recently, the isospin dependent elementary two-body $NN\to N\Delta$ cross section, i.e., $\tilde{\sigma}_{NN\to N\Delta}^* $, was studied in the framework of relativistic Boltzmann-Uehling-Uhlenbeck (RBUU) microscopic transport theory by  Li and Li in Ref. \cite{QingfengLi2017}.
Their results showed the $\tilde{\sigma}^*_{NN\rightarrow N\Delta}$ has a sharp increment around threshold energy without considering the $\Delta$ mass distribution, and medium correction factor $R=\tilde{\sigma}^*_{NN\rightarrow N\Delta}/\sigma^{\text{free}}_{NN\rightarrow N\Delta}$ obviously depends on the isospin channels of $NN\rightarrow N\Delta$, i.e., $pp\rightarrow n\Delta^{++}$, $pp\rightarrow p\Delta^{+}$, $pn\rightarrow n\Delta^{+}$, $pn\rightarrow p\Delta^{0}$, $nn\rightarrow n\Delta^{0}$, and $nn\rightarrow p\Delta^{-}$, in isospin asymmetric nuclear matter.
As a short living resonance, the $\Delta$ subsequently decays into nucleon and pion, and the measured cross section for $NN\rightarrow N\Delta$ is the elementary two-body cross section averaged over the mass distribution of $\Delta$ resonance, and thus medium correction factor $R$ including the effects from the mass distribution of $\Delta$ is worth to investigate.
Furthermore, the scalar and vector self-energies of incoming and outgoing particles are different in the $NN\rightarrow N\Delta$ process in isospin asymmetric nuclear matter, which named as threshold energy effects in $\Delta$ production\cite{Ferini2005,Song2015,ZhenZhang2017}. In our previous work \cite{Cui2018}, this effect on the in-medium $NN\rightarrow N\Delta$ cross section is analyzed. Our results confirm the isospin splitting of $R$ near the threshold energy in isospin asymmetric nuclear matter, but the splitting magnitude tends to vanish when the beam energy is above the 1.0 GeV.

In this paper, we study the in-medium $NN\rightarrow N\Delta$ cross sections and their differential cross sections under the three effective Lagrangian parameters in the isospin asymmetric nuclear matter, i.e., NL$\rho\delta$, DDME$\delta$ and DDRH$\rho\delta$, for further understanding the relation between the in-medium $NN\to N\Delta$ cross section and the nuclear matter parameters. 
The effective Lagrangian and the model of the in-medium $NN\rightarrow N\Delta$ cross section are briefly described in Sec.~\ref{model}.
In Sec.~\ref{xs}, we discuss the results of isospin dependent in-medium $NN\rightarrow N\Delta$ cross sections in different effective Lagrangian and analyze its relation to the effective mass, and briefly discuss its dependence on the slope of symmetry energy in the theoretical framework we used.  And a summary is given in Sec.~\ref{summary}.

\section{The Model}
\label{model}
\subsection{Effective Lagrangian and nuclear matter properties}
For the calculation of the in-medium $NN\rightarrow N\Delta$ cross section in isospin asymmetric nuclear matter, we use the one-boson exchange model with the relativistic Lagrangian including nucleon and $\Delta$ ($\Delta$ is the  Rarita-Schwinger spinor of spin-3/2 \cite{Huber1994,Machleidt1987,Benmerrouche1989}) which are coupled to $\sigma$, $\omega$, $\rho$, $\delta$, and $\pi$ mesons. Different from the work in Ref.~\cite{Larionov2003}, we include the isovector mesons $\rho$ and $\delta$ in order to describe the isospin asymmetric nuclear matter and isospin dependent in-medium $NN\rightarrow N\Delta$ cross section. The Lagrangian we used is as follows:
\begin{equation}
\label{Lag}	
\mathcal{L}=\mathcal{L}_I+\mathcal{L}_F,
\end{equation}
where $\mathcal{L}_F$ is
\begin{eqnarray}
\label{lag_f}
\mathcal{L}_{F}=&&\bar{\Psi}[i\gamma_{\mu}\partial^{\mu}-m_{N}]\Psi+\bar{\Delta}_{\lambda}[i\gamma_{\mu}\partial^{\mu}-m_{\Delta}]\Delta^{\lambda}\\
&&+\frac{1}{2}\left(\partial_{\mu}\sigma\partial^{\mu}\sigma-m_{\sigma}^2\sigma^2\right)-U(\sigma)\nonumber\\
&&-\frac{1}{4}\omega_{\mu\nu}\omega^{\mu\nu}+\frac{1}{2}m^{2}_{\omega}\omega_{\mu}\omega^{\mu}\nonumber\\
&&+\frac{1}{2}\left(\partial_{\mu}\bm{\pi}\partial^{\mu}\bm{\pi}-m^{2}_{\pi}\bm{\pi}^{2}\right)-\frac{1}{4}\bm{\rho}_{\mu\nu}\bm{\rho}^{\mu\nu}+\frac{1}{2}m^{2}_{\rho}\bm{\rho}_{\mu}\bm{\rho}^{\mu}\nonumber\\
&&+\frac{1}{2}\left(\partial_{\mu}\bm{\delta}\partial^{\mu}\bm{\delta}-m^{2}_{\delta}\bm{\delta}^{2}\right), \nonumber
\end{eqnarray}
$U(\sigma)$ is the nonlinear potential of $\sigma$ field,
\begin{eqnarray}
U(\sigma)=\left\{
\begin{array}{cc}
\frac{1}{3}g_{2}\sigma^{3}+\frac{1}{4}g_{3}\sigma^{4}& \text{NL}\rho\delta \\
0 & \text{DDME}\delta,  \text{DDRH}\rho\delta\\
\end{array} \right.
\end{eqnarray}
 $\mathcal{L}_I$ is
\begin{eqnarray}
\label{lag_i}
\mathcal{L}_I&=&\mathcal{L}_{NN}+\mathcal{L}_{\Delta \Delta}+\mathcal{L}_{N\Delta}\nonumber\\
&=&\Gamma_{\sigma NN}\bar{\Psi}\Psi\sigma-\Gamma_{\omega NN}\bar{\Psi}\gamma_{\mu}\Psi\omega^{\mu}-\Gamma_{\rho NN}\bar{\Psi}\gamma_{\mu}\bm{\tau} \cdot\Psi\bm{\rho}^{\mu}\nonumber\\
&&+\frac{g_{\pi NN}}{m_{\pi}}\bar{\Psi}\gamma_{\mu}\gamma_{5}\bm{\tau} \cdot\Psi\partial^{\mu}\bm{\pi}+\Gamma_{\delta NN}\bar{\Psi}\bm{\tau} \cdot\Psi\bm{\delta}\nonumber\\
&&+\Gamma_{\sigma \Delta \Delta}\bar{\Delta}_{\mu}\Delta^{\mu}\sigma-\Gamma_{\omega \Delta \Delta}\bar{\Delta}_{\mu}\gamma_{\nu}\Delta^{\mu}\omega^{\nu} \nonumber\\
&&-\Gamma_{\rho \Delta\Delta}\bar{\Delta}_{\mu}\gamma_{\nu}\bm{\mbox{T}} \cdot\Delta^{\mu}\bm{\rho}^{\nu}+\frac{g_{\pi \Delta\Delta}}{m_{\pi}}\bar{\Delta}_{\mu}\gamma_{\nu}\gamma_{5}\bm{\mbox{T}} \cdot\Delta^{\mu}\partial^{\nu}\bm{\pi}\nonumber\\
&&+\Gamma_{\delta \Delta\Delta}\bar{\Delta}_{\mu}\bm{\mbox{T}} \cdot\Delta^{\mu}\bm{\delta}+\frac{g_{\pi N\Delta}}{m_{\pi}}\bar{\Delta}_{\mu}\bm{\mathcal{T}}\cdot \Psi\partial^{\mu}\bm{\pi}\nonumber\\
&&+\frac{ig_{\rho N\Delta}}{m_{\rho}}\bar{\Delta}_{\mu}\gamma_{\nu}\gamma_{5}\bm{\mathcal{T}}\cdot \Psi\left(\partial^{\nu}\bm{\rho}^{\mu}-\partial^{\mu}\bm{\rho}^{\nu}\right)+h.c. ~
\end{eqnarray}
 $\omega_{\mu\nu}$ and $\bm{\rho}_{\mu\nu}$ in Eq.(\ref{lag_f}) are defined by $\partial_{\mu}\omega_{\nu}-\partial_{\nu}\omega_{\mu}$ and $\partial_{\mu}\bm{\rho}_{\nu}-\partial_{\nu}\bm{\rho}_{\mu}$, respectively. Here $\bm{\tau}$ and $\mathbf{T}$ are the isospin matrices of nucleon and $\Delta$ \cite{Machleidt1987,Benmerrouche1989}, and $\bm{\mathcal{T}}$ is the isospin transition matrix between the isospin 1/2 and the 3/2 fields \cite{Huber1994}. $\Gamma_{m NN}$  is meson-nucleon  coupling constant
\begin{eqnarray}
\Gamma_{m NN}=\left\{
\begin{array}{cc}
g_{m NN}& \text{NL}\rho\delta \\
g_{m NN}(\rho_B) & \text{DDME}\delta, \text{DDRH}\rho\delta\\
\end{array} \right.
\end{eqnarray}
the values of $\Gamma_{mNN}$ are listed in Table I.

For the coupling constants $\Gamma_{m\Delta\Delta}$, $m=\sigma, \omega, \rho, \delta$, we simply take them to be equal to the meson-nucleon-nucleon coupling, i.e., $\Gamma_{m\Delta\Delta}=\Gamma_{mNN}$, as the same as transport models calculations \cite{Song2015,QingfengLi2017,Larionov2003}. The coupling constant $g_{\pi N\Delta}$ is indispensable for describing the $NN\to N\Delta$ cross section, and it is determined by analyzing the $\Delta$-isobar decay width from Ref.\cite{Dmitriev1986}.
But there is no contribution on EOS from $\pi$ meson in the relativistic mean field without Fock term. Concerning the coupling constant $g_{\rho N\Delta}$, we use $g_{\rho N\Delta}\approx\frac{\sqrt{3}}{2} \Gamma_{\rho NN} \frac{m_{\rho}}{m_N}$ which are derived from the static quark model \cite{Engel90,Huber1994}. 
\begin{table}[htbp]
\begin{center}
\caption{The parameters used in the effective Lagrangian, $g_{\pi NN}$=1.008, $g_{\pi N\Delta}$=2.202, $m_{\pi}$=138, $m_{N}$=939, $m_{0,\Delta}$=1232 (all masses are in MeV), $g_{2}/g_{\sigma NN}^3$=0.03302 fm$^{-1}$(NL$\rho\delta$), $g_{3}/g_{\sigma NN}^4$=-0.00483 (NL$\rho\delta$), $\Lambda_{\pi NN}$=1000MeV. The coupling constants $\Gamma_{mNN}$ and $g_{mN\Delta}$ are dimensionless.}\label{table-para}
\begin{threeparttable}
\begin{tabular}{c | c | c | c}
  \hline
  \hline
     & NL$\rho\delta$-$\Delta$ & DDME$\delta$-$\Delta$\tnote{a} & DDRH$\rho\delta$-$\Delta$\tnote{a} \\
    \hline
    $m_{\sigma}$ (MeV) & 550 & 566 & 550  \\
    $m_{\omega}$ (MeV)& 783 &  783 & 783 \\
    $m_{\rho}$ (MeV)& 770 & 769 & 763 \\
    $m_{\delta}$ (MeV) & 980 &983 & 980\\
   $\Gamma_{\sigma NN}$  &8.9679 & 10.3313 & 10.7286 \\
   $\Gamma_{\omega NN}$  &9.2408 &  12.2905 & 13.2902 \\
   $\Gamma_{\rho NN}$  &6.9256 &  6.3117 & 5.8284 \\
   $\Gamma_{\delta NN}$  &7.8525 &  7.1515 &  7.6009\\
   $\Lambda_{\pi N\Delta}$ (MeV) &410 &  416 & 417 \\
   $\Lambda_{\rho NN}$ (MeV) &1000 &  650 & 580 \\
   \hline
   \hline
   $E/A$ (MeV) & $-16.00$ & $-16.12$ & $-16.25$  \\
$\rho_{0}$ (fm$^{-3}$)& 0.160 &0.152& 0.153 \\
   $K_0$(MeV)& 240.0 & 219.1 & 240.2 \\
    $S_{0}$(MeV) & 30.60 & 32.35  & 25.34 \\
   $L$(MeV) &101.46 &  52.85 & 45.33 \\
   $m^{*}_{N}/m_{N}$ &0.75 &  0.609 & 0.55 \\
 $m^{*}_{\Delta}/m_{\Delta}$ & 0.809 & 0.702  & 0.661 \\
   $\Delta m^*_N$ \tnote{b}&0.0312 &  0.0236 & 0.0265 \\
   $\Delta m^*_\Delta$\tnote{b} & 0.0079  & 0.0060  & 0.0068  \\

   \hline
   \hline	 		 		 	 	 			 	 	
\end{tabular}
\begin{tablenotes}\small
\item[a] The density dependent coupling constants of DDME$\delta$-$\Delta$ and DDRH$\rho\delta$-$\Delta$  are for $\rho_B=\rho_0$.
\item[b] Here $\Delta m^*_N=\frac{m^{*}_{p}-m^{*}_{n}}{m_{N}}$ and $\Delta m^*_\Delta=\frac{m^{*}_{\Delta^{++}}-m^{*}_{\Delta^+}}{m_{\Delta}}$.
\end{tablenotes}
\end{threeparttable}
\end{center}

\end{table}


The coupling constants of nucleon to $\sigma$, $\omega$, $\rho$, and $\delta$ mesons are important for prediction of the in-medium  $NN\rightarrow N\Delta$ cross section as well as for the EOS.
In this work, we select three parameter sets, i.e., NL$\rho\delta$, DDME$\delta$ and DDRH$\rho\delta$ from five alternative sets \cite{Hofmann2001, Liu2002, Gaitanos2004, Gogelein2008, Roca2011} which contain $\sigma$, $\omega$, $\rho$ and $\delta$, and the compressibility is in reasonable region, i.e., $K_0 = 230\pm 40$ MeV as in \cite{Dutra14}. For the NL$\rho\delta$ parameter set, $U(\sigma)$ includes the nonlinear $\sigma$ self-interaction which can reproduce reasonable values of the incompressibility and nucleon effective mass by adding two additional free parameters, but it can also be realized by adopting the density dependent coupling constants in DDME$\delta$ \cite{Roca2011} and DDRH$\rho\delta$ \cite{Gaitanos2004}.
Since we included $\Delta$ degree in the effective Lagrangian, we named them as  NL$\rho\delta$-$\Delta$, DDME$\delta$-$\Delta$, DDRH$\rho\delta$-$\Delta$ in the following text to distinguish them from their original name of parameter sets in the relativistic mean field model (RMF).

In the rest nuclear matter, the effective momentum can be written as $\textbf{p}_i^*=\textbf{p}_i$ since the spatial components of vector field vanish, i.e., $\mathbf{\Sigma}=0$. Thus, in the mean field approach, the effective energy reads as
\begin{equation}
p_i^{*0}=p^{0}_{i}-\Sigma^{0}_{i},
\end{equation}
and
\begin{equation}
\Sigma^{0}_{i}=\Gamma_{\omega NN}\bar{\omega}^{0}+\Gamma_{\rho NN}t_{3,i}\bar{\rho}^{0}_3.
\end{equation}
Here $t_{3,i}$ is the third component of the isospin of the nucleon and $\Delta$, and i=n, p, $\Delta^{++}$, $\Delta^{+}$, $\Delta^{0}$, $\Delta^{-}$, where $t_{3,n}=-1$, $t_{3,p}=1$, $t_{3,\Delta^{++}}=1$, $t_{3,\Delta^{+}}=\frac{1}{3}$, $t_{3,\Delta^{0}}=-\frac{1}{3}$, $t_{3,\Delta^{-}}=-1$, and $\bar{\rho}^{0}_3=\frac{\Gamma_{\rho NN}}{m^2_\rho}(\rho_{p}-\rho_{n}$).
The Dirac effective masses of nucleon and $\Delta$ read as:
\begin{equation}
m^{*}_{i}=m_{i}+\Sigma^{S}_{i},
\label{eq:efmnd}
\end{equation}
where
\begin{equation}
\Sigma^{S}_{i}=-\Gamma_{\sigma NN}\bar{\sigma}- \Gamma_{\delta NN}t_{3,i}\bar{\delta}_3,
\label{eq:efmnd2}
\end{equation}
and $\bar{\delta}_3=\frac{\Gamma_{\delta NN}}{m^2_{\delta}}(\rho^{S}_p-\rho^{S}_n)$. %

The density dependent of symmetry energy is:
\begin{eqnarray}
\label{symmetry}
S(\rho_B)=&&\frac{k^{2}_{F}}{6E^{*}_{F}}+\frac{\Gamma^2_{\rho NN}}{2m^2_{\rho}}\rho_{B}\\
         &&-\frac{1}{2}\frac{\Gamma^2_{\delta NN}}{m^2_{\delta}}\frac{m^{*2}_{N}\rho_B}{E^{*2}_{F}(1+\frac{\Gamma^2_{\delta NN}}{m^2_{\delta}}A(k_{F},m^{*}_{N}))}.\nonumber
\end{eqnarray}
which depends on the effective mass, coupling constant of $\Gamma_{\rho NN}$, and $\Gamma_{\delta NN}$. The slope of  symmetry energy $L$ is:
\begin{equation}
\label{Le}
L = 3\rho_0\frac{dS(\rho_B)}{d\rho_B}\mid_{\rho_{B}=\rho_0}=L^{\mathrm{kin}}+L^{\rho}+L^{\delta}
\end{equation}
where
\begin{equation}
\label{Lk}
L^{\mathrm{kin}}=\frac{k^{2}_{F}}{6E^{*}_{F}}(2-\frac{k^{2}_{F}}{E^{*2}_{F}}-\frac{3m^{*2}_{N}}{E^{*2}_{F}}\frac{\rho_{0}}{m^{*}_{N}}\frac{\partial m^{*}_{N}}{\partial \rho_{B}})
\end{equation}
\begin{equation}
\label{Lr}
L^{\rho}=\frac{\Gamma^2_{\rho NN}}{2m^2_{\rho}}\rho_{0}(3+6\frac{\rho_0}{\Gamma_{\rho NN} }\frac{\partial \Gamma_{\rho NN}}{\partial \rho_B})
\end{equation}
\begin{eqnarray}
\label{Ld}
L^{\delta}=&&-\frac{1}{2}\frac{\Gamma^2_{\delta NN}}{m^2_{\delta}}\frac{m^{*2}_{N}\rho_0}{E^{*2}_{F}(1+(\frac{\Gamma_{\delta NN}}{m_{\delta}})^2A)}\nonumber\\  &&\times\{3+6\frac{\rho_0}{\Gamma_{\delta NN} }\frac{\partial \Gamma_{\delta NN}}{\partial \rho_B}-\frac{2k^{2}_{F}}{E^{*2}_{F}}\nonumber\\
&&+6(1-\frac{m^{*2}_{N}}{E^{*2}_{F}})\frac{\rho_{0}}{m^{*}_{N}}\frac{\partial m^{*}_{N}}{\partial \rho_{B}}\nonumber\\
&&-3\frac{\Gamma^2_{\delta NN}}{m^2_{\delta}}\frac{1}{1+\frac{\Gamma^2_{\delta NN}}{m^2_{\delta}}A}\nonumber\\
&&\times[2A(\frac{\rho_0}{\Gamma_{\delta NN} }\frac{\partial \Gamma_{\delta NN}}{\partial \rho_B}+\frac{\rho_{0}}{m^{*}_{N}}\frac{\partial m^{*}_{N}}{\partial \rho_{B}})\nonumber\\
&&+\rho_0\frac{k^{2}_{F}}{E^{*3}_{F}}(1-3\frac{\rho_{0}}{m^{*}_{N}}\frac{\partial m^{*}_{N}}{\partial \rho_{B}}]\}
\end{eqnarray}
with $E^*_{F}=\sqrt{k^2_{F}+m^{*2}_{N}}$ and
\begin{equation}
\label{factora}
A = \frac{2}{\pi^2}\int^{k_{F}}_0\frac{k^4dk}{(k^2+m^{*2}_N)^{3/2}}.
\end{equation}

The corresponding nuclear matter parameters at normal density are listed in the lower part of Table~\ref{table-para}, where the NL$\rho\delta$-$\Delta$ predicts the largest slope of symmetry energy $L$, effective mass $m^*$, and effective mass splitting $\Delta m^*_{N}=(m^*_p-m^*_n)/m_N$ and $\Delta m^*_{\Delta}=(m^*_{\Delta^{++}}-m^*_{\Delta^{+}})/m_{\Delta}$, among  these three parameter sets at normal density.
For the symmetry energy coefficient $S_{0}$,  the DDRH$\rho\delta$-$\Delta$ predicts the smallest value and DDME$\delta$-$\Delta$ predicts the largest value among the three selected parameter sets. 
Among the three selected parameter sets, the larger $L$ corresponds to larger $m^*$. 

\begin{figure}[htbp]
\begin{center}
    \includegraphics[scale=0.48]{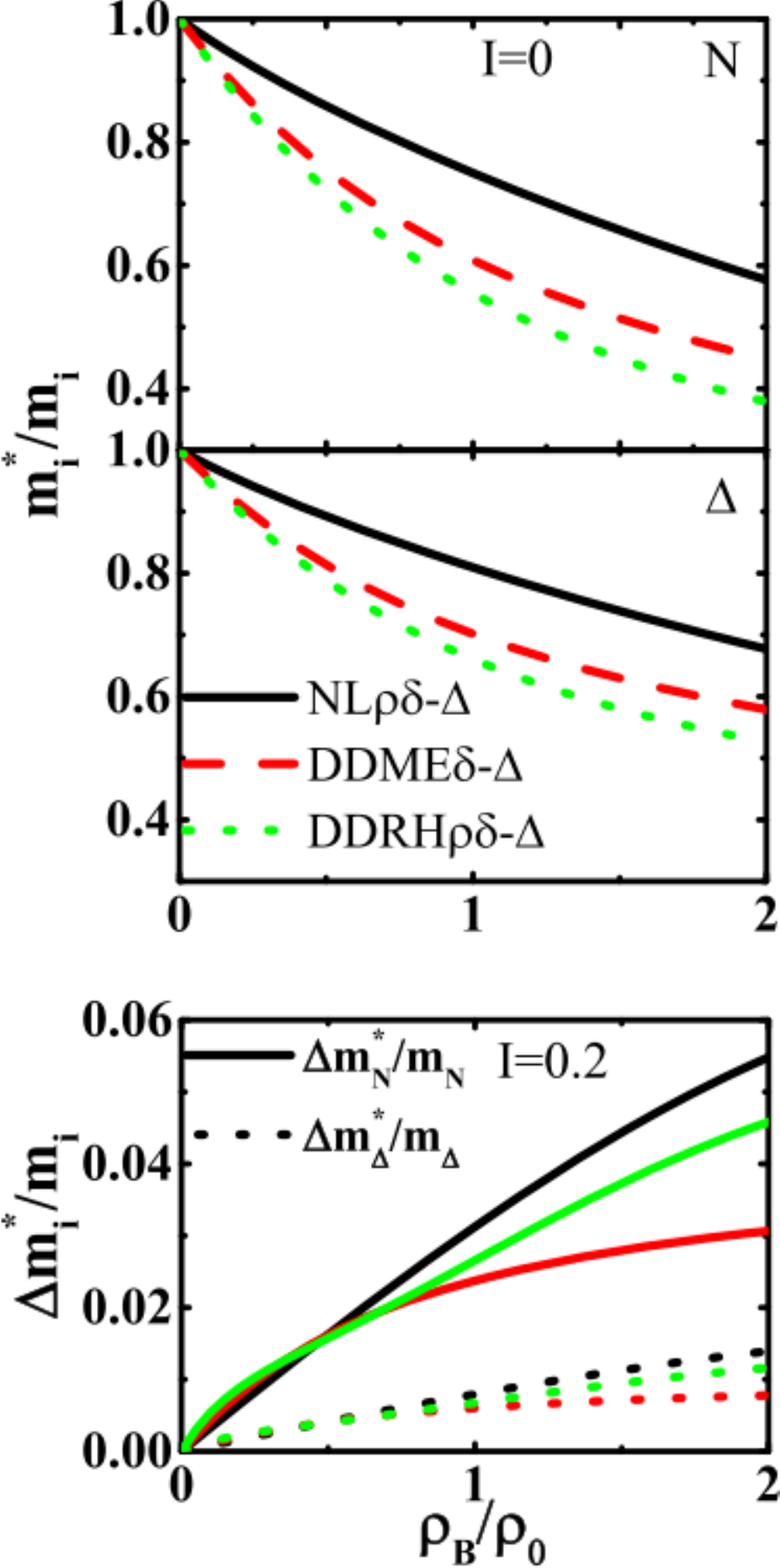}
    \caption{(Color online) (a) and (b)  the effective masses of nucleon   and the effective pole masses of $\Delta$ as a function of $\rho_B/\rho_0$ in symmetric nuclear matter. (c)  the effective masses splitting as a function of density for nucleons and $\Delta$s at $I$=0.2.  }\label{fig-efm}
\end{center}
\end{figure}

In Fig.~\ref{fig-efm}, we present the Dirac effective masses as functions of density for nucleon and $\Delta$ in symmetric nuclear matter, the black solid lines, the red dashed, and the green dotted lines are the results for  NL$\rho\delta$-$\Delta$, DDME$\delta$-$\Delta$ and DDRH$\rho\delta$-$\Delta$ respectively.
The upper panel is the effective masses for nucleons and the middle panel is for the effective $\Delta$ pole masses.  Among the selected parameter sets, the NL$\rho\delta$-$\Delta$ has the largest effective mass, while the DDRH$\rho\delta$-$\Delta$ has the smallest value.
In symmetric nuclear matter,  $m_N^*/m_N$=0.75, $m_N^*/m_N$=0.609, and $m_N^*/m_N$=0.55 for  NL$\rho\delta$-$\Delta$, DDME$\delta$-$\Delta$ and DDRH$\rho\delta$-$\Delta$ at saturation density, respectively.
In the neutron-rich matter, the effective masses of nucleons and $\Delta$'s are split due to the contributions from isovector-scalar $\delta$ meson. There is $m_p^*>m_n^*$, $m_{0,\Delta^{++}}^*>m_{0,\Delta^{+}}^*>m_{0,\Delta^{0}}^*>m_{0,\Delta^{-}}^*$ in the neutron-rich matter. The splitting magnitude of the effective masses for nucleons and $\Delta$s depends on the coupling constant $\Gamma_{\delta NN}$($\Gamma_{\delta\Delta\Delta}$) and $\bar{\delta}_3$ in Eq.~\ref{eq:efmnd2}. Here, we define the splitting magnitude of the effective mass as, $\Delta m^{*}_N/m_{N}=(m^*_p-m^*_n)/m$ and $\Delta m^*_\Delta/m_\Delta=(m^*_{\Delta^{++}}-m^*_{\Delta^+})/m_\Delta$. As shown in the bottom panel of Fig.~\ref{fig-efm}, the NL$\rho\delta$-$\Delta$ gives the largest effective mass splitting above normal density, but the two other parameter sets DDME$\delta$-$\Delta$ and DDRH$\rho\delta$-$\Delta$ predict comparatively small effective mass splitting since the strength of $\Gamma_{\delta NN}$($\Gamma_{\delta\Delta\Delta}$) decrease with the density.


\subsection{In-medium $NN\rightarrow N\Delta$ cross section}

In quasiparticle approximation \cite{Baym76}, the in-medium cross sections are introduced via the replacement of the vacuum plane waves of the initial and final particles by the plane waves obtained by solution of the nucleon and $\Delta$ equation  of motion with scalar and vector fields. In detail, the matrix elements $\mathcal{M}^*$ for the inelastic scattering process $NN\rightarrow N\Delta$ are obtained by replacing the nucleon and $\Delta$ masses and momenta in free space with their effective masses and kinetic momenta \cite{Larionov2003}, i.e., $m \to m^*$ and $p^{\mu}\to p^{* \mu}$.
All the calculations performed in this work are in the center-of-mass frame of colliding particles, it  coincides with the nuclear matter rest frame, where the spatial components of the vector field vanish \cite{Larionov2003}.

\begin{figure}[htbp]
\begin{center}
    \includegraphics[scale=0.1]{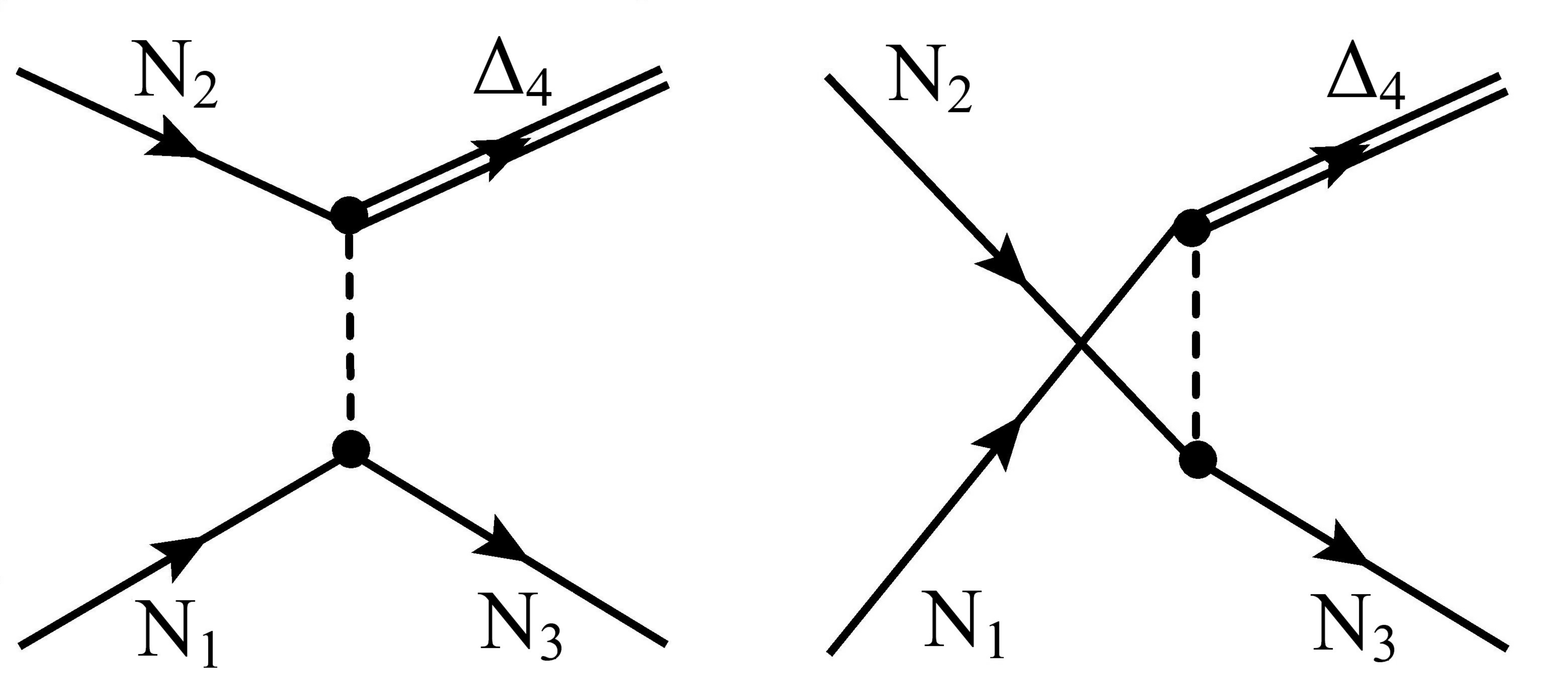}
    \caption{The left diagram is the direct term, and the right one is the exchange term.}\label{fig0}
\end{center}
\end{figure}

The Feynmann diagrams corresponding to the inelastic-scattering $NN\rightarrow N\Delta$ processes are shown in Fig. ~\ref{fig0}, which include the direct and exchange processes. The $\mathcal{M}^*$-matrix for the interaction Lagrangian Eq.~(\ref{lag_i}) can be written by the standard procedure \cite{Huber1994},

\begin{equation}
\mathcal{M}^*=\mathcal{M}_d^{*\pi}-\mathcal{M}_e^{*\pi}+\mathcal{M}_d^{*\rho}-\mathcal{M}_e^{*\rho}
\end{equation}
where
\begin{eqnarray}
\mathcal{M}_d^{*\pi}&=&-i\frac{g_{\pi NN} g_{\pi N\Delta} I_d  }{ m_{\pi}^{2}( Q^{*2}_{d}- m_{\pi}^{2})}[\bar{\Psi}(p_3^* ) \gamma_{\mu}\gamma_5  Q_{d}^{*\mu}  \Psi(p_1^*)]\nonumber\\
&&\times[\bar{\Delta}_{\nu} (p_4^* ) Q_d^{*\nu}  \Psi(p_2^* )]\\
\mathcal{M}_d^{*\rho}&=&i\frac{\Gamma_{\rho NN} g_{\rho N\Delta}I_d}{m_{\rho} }[\bar{\Psi}(p_3^* ) \gamma_{\mu} \Psi(p_1^* )]\\\nonumber
&&\times\frac{g^{\mu\tau}-Q_d^{*\mu} Q_d^{*\tau}/m^{2}_{\rho}}{Q_d^{*2}-m^{2}_{\rho}}\\\nonumber
&&\times[\bar{\Delta}_{\sigma} (p_4^* ) \gamma_{\lambda} \gamma_{5} (Q_d^{*\lambda} \delta_{\sigma\tau}-Q_d^{*\sigma} \delta_{\lambda\tau}) \Psi(p_2^* )]
\end{eqnarray}


The upper index in $\mathcal{M}^{*\text{meson}}_{d,e}$ refers to the exchanged boson, the lower index to the direct or exchange process. $Q_{d}^{*\mu}=p_{3}^{*\mu}-p_{1}^{*\mu}$ for the direct term, the exchange term $\mathcal{M}^*_e$ is obtained by $p_{1}^{*\mu}\longleftrightarrow p_{2}^{*\mu}$ and $Q_{e}^{*\mu}=p_{3}^{*\mu}-p_{2}^{*\mu}$. The isospin factors $I_d$, $I_e$  can be found in the Ref.~\cite{Huber1994}.

The in-medium $NN\rightarrow N\Delta$ cross section is the in-medium elementary two-body cross section averaged over the mass of $\Delta$ by considering the $\Delta$ as the short-living resonance, and it can be written as:
\begin{eqnarray}
&&\sigma^*_{NN\rightarrow N\Delta}=\int_{m^*_{\Delta,\text{min}}}^{m^*_{\Delta,\text{max}}} dm^*_{\Delta}f(m^*_{\Delta})\tilde{\sigma}^*(m^*_{\Delta})
\label{eq:xsnd1}
\end{eqnarray}
$\tilde{\sigma}^*( m^*_{\Delta})$ is the in-medium elementary two-body cross section. In the center-of-mass frame of colliding nucleons, it reads
\begin{eqnarray}
\label{eq:xsnd2}
\tilde{\sigma}^*( m^*_{\Delta})
&=&\frac{1}{4F^*}\int
\frac{d^3 \textbf{p}_3^*}{(2\pi)^3 2E_3^* }  \frac{d^3 \textbf{p}_4^*}{(2\pi)^3 2E_4^* }\\\nonumber
&&\times(2\pi)^4\delta^{4}(p_1+p_2-p_3-p_4)\overline{|\mathcal{M}^*|^2}\\\nonumber
&=&\frac{1}{64\pi^2}\int \frac{|\textbf{p}^{*}_{\text{out, c.m.}}|}{\sqrt{s^*_{\text{in}}}\sqrt{s^*_{\text{out}}}|\textbf{p}^{*}_{\text{in, c.m.}}|} \overline{|\mathcal{M}^*|^2}  d\Omega,
\end{eqnarray}
where $\textbf{p}^{*}_{\text{in, c.m.}}$ and $\textbf{p}^{*}_{\text{out, c.m.}}$ are the momenta of incoming (1 and 2) and outgoing particles (3 and 4), respectively. $F^*=\sqrt{(p^*_{1}p^*_{2})^2-p^{*2}_{1}p^{*2}_{2}}=\sqrt{s^*_{\text{in}}}|\textbf{p}^{*}_{\text{in, c.m.}}|$ is the invariant flux factor, $s^*_{\text{in}}=(p^*_1+p^*_2)^2$, and $s^*_{\text{out}}=(p^*_3+p^*_4)^2$.
Here $\overline{|\mathcal{M}^*|^2}=\frac{1}{(2s_{1}+1)(2s_2+1)}\sum\limits_{s_{1}s_{2}s_{3}s_{4}}|\mathcal{M}^*|^2$ is,
\begin{eqnarray}
&&\sum\limits_{s_{1}s_{2}s_{3}s_{4}}|\mathcal{M}^*|^2 \nonumber\\
&&=\sum\limits_{s_{1}s_{2}s_{3}s_{4}} \{ |\mathcal{M}_d^{*\pi}|^2-\mathcal{M}_d^{*\pi \dagger}\mathcal{M}_e^{*\pi}-\mathcal{M}_e^{*\pi \dagger}\mathcal{M}_d^{*\pi}+|\mathcal{M}_e^{*\pi}|^2\nonumber\\
&& +|\mathcal{M}_d^{*\rho}|^2-\mathcal{M}_d^{*\rho \dagger}\mathcal{M}_e^{*\rho}-\mathcal{M}_e^{*\rho \dagger}\mathcal{M}_d^{*\rho}+|\mathcal{M}_e^{*\rho}|^2 \nonumber\\
&&+\mathcal{M}_d^{*\pi \dagger}\mathcal{M}_d^{*\rho}-\mathcal{M}_d^{*\pi \dagger}\mathcal{M}_e^{*\rho}-\mathcal{M}_e^{*\pi \dagger}\mathcal{M}_d^{*\rho}+\mathcal{M}_e^{*\pi \dagger}\mathcal{M}_e^{*\rho} \nonumber\\
&&+\mathcal{M}_d^{*\rho \dagger}\mathcal{M}_d^{*\pi}-\mathcal{M}_d^{*\rho \dagger}\mathcal{M}_e^{*\pi}-\mathcal{M}_e^{*\rho \dagger}\mathcal{M}_d^{*\pi}+\mathcal{M}_e^{*\rho \dagger}\mathcal{M}_e^{*\pi}\}. \nonumber\\
\end{eqnarray}
All the terms are calculated by  Mathematics with the packages of ``High Energy Physics" \cite{math}.
Here, we only show the direct term as an example for $\pi$ mesons, i.e., $\sum\limits_{s_{1}s_{2}s_{3}s_{4}} |\mathcal{M}_d^{*\pi}|^2$:
\begin{eqnarray}
\label{eq:xsnd3}
&&\sum\limits_{s_{1}s_{2}s_{3}s_{4}} |\mathcal{M}_d^{*\pi}|^2 =\left(\frac{g_{\pi NN} g_{\pi N\Delta} I_d  }{ m_{\pi}^{2}( Q^{*2}_{d}- m_{\pi}^{2})}\right)^2\nonumber\\
&&\times\sum\limits_{s_{1}s_{2}s_{3}s_{4}}[\Psi(p_1^*)\bar{\Psi}(p_1^* ) \gamma_{\mu}\gamma_5  Q_{d}^{*\mu}  \Psi(p_3^*)\bar{\Psi}(p_3^* )\gamma_{\sigma}\gamma_5  Q_{d}^{*\sigma}]\nonumber\\
&&\times[\Psi(p_2^* )\bar{\Psi} (p_2^* ) Q_d^{*\nu}  \Delta_{\nu} (p_4^* )\bar{\Delta}_{\tau} (p_4^* ) Q_d^{*\tau}]\nonumber\\
&&=\left(\frac{g_{\pi NN} g_{\pi N\Delta} I_d  }{ m_{\pi}^{2}( t^{*}- m_{\pi}^{2})}\right)^2\nonumber\\
&&\times\frac{2 (m^*_{N_{1}} + m^{*}_{N_{3}})^2((m^{*}_{N_{1}} - m^{*}_{N_{3}})^2 - t^*)}{3 m^{*2}_{\Delta_{4}}}\nonumber\\
&&\times\left((m^{*}_{\Delta_{4}}-m^{*}_{N_{2}})^2 - t^*\right) \left((m^{*}_{N_{2}} +  m^{*}_{\Delta_{4}})^2 - t^*\right)^2
\end{eqnarray}
where $t=Q^{*2}_{d}$, for $|\mathcal{M}_e^{*\pi}|^2$ is $N_{1}\leftrightarrow N_{2}$. In Eq. ~(\ref{eq:xsnd2}), one should notice that the crucial requirement of two-body collisions is the energy-momentum conservation in terms of incoming and outgoing canonical momenta ($p^\mu_{1,2}$, $p^\mu_{3,4}$), i.e., $\delta^4(p_1+p_2-p_3-p_4)$. In the language of kinetic momentum, the energy-momentum conservation $p^\mu_{1}+p^\mu_{2}= p^\mu_{3}+p^\mu_{4}$ can be expressed as $p^{*\mu}_{1}+\Sigma^{*\mu}_{1}+ p^{*\mu}_{2}+\Sigma^{*\mu}_{2}= p^{*\mu}_{3}+\Sigma^{*\mu}_{3}+p^{*\mu}_{4}+\Sigma^{*\mu}_{4}$, $p^{*\mu}_{1}+ p^{*\mu}_{2}= p^{*\mu}_{3}+p^{*\mu}_{4}-\Delta\Sigma^{\mu}$, here $\Delta\Sigma^{\mu}=\Sigma^{\mu}_{1}+\Sigma^{\mu}_{2}-\Sigma^{\mu}_{3}-\Sigma^{\mu}_{4}$ is the kinetic momentum change between the initial and final states, and the effective energy changes are expressed as $\Delta \Sigma^0=\Sigma^{0}_{1}+\Sigma^{0}_{2}-\Sigma^{0}_{3}-\Sigma^{0}_{4}$, which is as same as in the formula in Ref.\cite{ZhenZhang2018}. The similar issue also exists in the calculation of $m^{*}_{\text{min}}$,  $m^{*}_{\text{max}}$ and $\Gamma(m^{*}_{\Delta})$ in the following.

The $m^*_{\Delta,\text{min}}$ in the formula of the cross section is determined by the $\Delta \rightarrow N+ \pi$ in isospin asymmetric nuclear matter as in  Refs. \cite{ZhenZhang2017,Cui2018} when both $N$ and $\pi$ are at rest, and the modification of scalar and vector self-energies in this isospin exchange process should also be considered. Thus, $m^*_{\Delta,\text{min}}=m^*_{N}+\Sigma^{0}_{N}+m^*_\pi+\Pi_P(\omega, \mathbf{q})-\Sigma^{0}_{\Delta}$=$m^*_{N}+m^*_\pi-\Delta\Sigma_d^0$, with $\Delta\Sigma^0_d=\Sigma^{0}_{N}+\Pi_P(\omega, \mathbf{q})-\Sigma^{0}_{\Delta}$. Considering  $m^*_{\pi}/m_{\pi}$ is less than $\sim10\%$ at normal density from the calculations by Kaiser and Weise \cite{Kaiser01}, we simply neglect the effect that the pions are affected by the nuclear mean field and take $m^*_{\pi}=m_{\pi}$ in this paper. Thus, we have $\Delta\Sigma_d^0=\Sigma_\Delta^0-\Sigma_{N}^0$.
The $m^*_{\Delta,\text{max}}$ is evaluated from $NN\to \Delta N$ for producing $N$ and $\Delta$ at rest, and it leads to
\begin{equation}
m^*_{\Delta,\text{max}}=\sqrt{s}-m^*_{N_{3}}-\Sigma^0_{N_{3}}-\Sigma^0_{\Delta_4}.
\end{equation}

The in-medium $\Delta$ mass distribution $f(m^*_\Delta)$ is another important ingredient of in-medium $NN\rightarrow N\Delta$ cross section for which proper energy conservation is also necessary since  $f(m^*_\Delta)$ is related to the $\Delta\rightarrow N+\pi$ process in isospin asymmetric nuclear matter. In this paper, the spectral function of  $\Delta$ is taken as in Ref. \cite{Larionov2003},
\begin{equation}
\label{eq:bt}
f(m^*_{\Delta})=\frac{2}{\pi}\frac{m^{* 2}_{\Delta}\Gamma(m^{*}_{\Delta})}{(m^{*2}_{0,\Delta}-m^{*2}_{\Delta})^2+m^{*2}_{\Delta}\Gamma^2(m^{*}_{\Delta}) }.
\end{equation}
Here, $m^*_{0,\Delta}$ is the effective pole mass of $\Delta$ and $\frac{2}{\pi}$ is the normalization factor. The decay width $\Gamma(m^*_\Delta)$ is taken as the parameterization form \cite{Larionov2003}
\begin{eqnarray}
\label{eq:gama}
\Gamma(m^{*}_{\Delta})&=&\Gamma_{0}\frac{q^{3}(m^{* }_{\Delta},m^*_N,m^*_\pi)}{q^{3}(m^{*}_{0,\Delta},m^*_N,m^*_\pi)}\\\nonumber
&&\times\frac{q^{3}(m^{*}_{0,\Delta},m^*_N,m^*_\pi)+\eta^2}{q^{3}(m^{* }_{\Delta},m^*_N,m^*_\pi)+\eta^2}\frac{m^{*}_{0,\Delta}}{m^{*}_{\Delta}},
\end{eqnarray}
where
\begin{eqnarray}
\label{eq:qm123}
&&q(m^{*}_\Delta,m^*_{N},m^*_\pi)=\\\nonumber
&&\sqrt{\frac{\left((m^*_\Delta+\Sigma^{0}_{\Delta}-\Sigma^{0}_{N})^2+m_{N}^{*2}-m_{\pi}^{* 2}\right)^2}
{4(m^{*}_\Delta+\Sigma^{0}_{\Delta}-\Sigma^{0}_{N})^2}-m_{N}^{*2}}
\end{eqnarray}
is the center-of-mass momentum of nucleon and pion from the decay of $\Delta$ in its rest frame. The factor of $(m^*_\Delta+\Sigma^{0}_{\Delta}-\Sigma^{0}_{N})$ in Eq.~(\ref{eq:qm123}) comes from properly considering the energy conservation in $\Delta\rightarrow N\pi$ process in the isospin asymmetric nuclear matter.
The coefficients of $\Gamma_0$=0.118 GeV and $\eta$=0.2 GeV/$c$ are used in the above parameterization formula.

\begin{figure}[htbp]
\begin{center}
\includegraphics[scale=0.52]{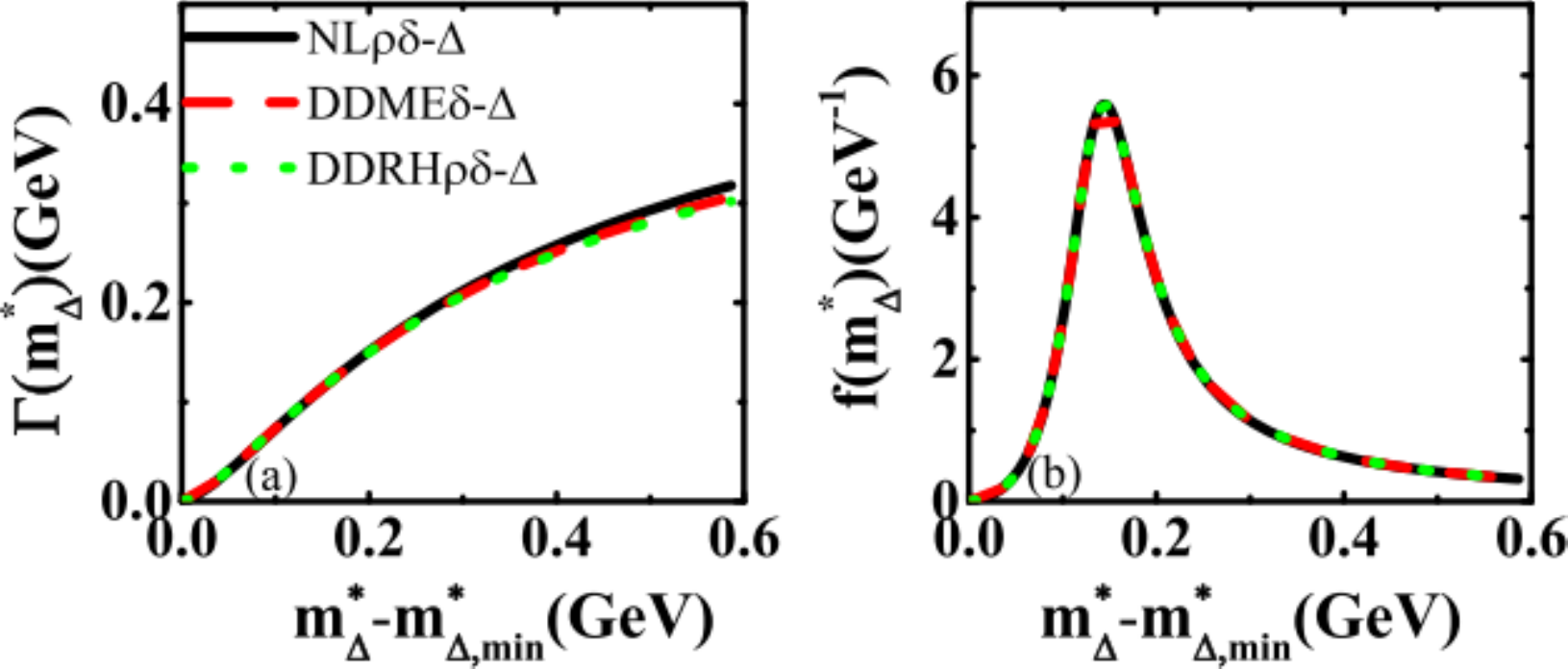}
\caption{(Color online) (a) $\Gamma(m^{*}_{\Delta})$ and (b) $f(m^{*}_{\Delta})$ as a function of $m^{* }_{\Delta}-m^{* }_{\Delta, \text{min}}$ at $\rho_B=\rho_0$ for symmetric nuclear matter $I=0$. The black, red, and green lines are the results for NL$\rho\delta$-$\Delta$, DDME$\delta$-$\Delta$ and DDRH$\rho\delta$-$\Delta$ respectively.}\label{fig4d}
\end{center}
\end{figure}

As an example, we present the decay width $\Gamma(m^{*}_{\Delta})$ and the $f(m^{*}_{\Delta})$ as a function of $m^{* }_{\Delta}-m^{* }_{\Delta, \text{min}}$ in Figure~\ref{fig4d} for symmetric nuclear matter $I=0$ and $\rho_B=\rho_0$ since their dependence on isospin asymmetry and density is negligible.
$m^{* }_{\Delta}-m^{* }_{\Delta, \text{min}}$ is used in order to compare the $\Gamma(m^{*}_{\Delta})$ and $f(m^{*}_{\Delta})$ in different parameter sets, because the $m^{* }_{\Delta, \text{min}}$ are different in different parameter sets, such as NL$\rho\delta$-$\Delta$ (black lines), DDME$\delta$-$\Delta$ (red lines), and DDRH$\rho\delta$-$\Delta$ (green lines).
Based on the Eq.~\ref{eq:qm123}, the values of $m^*_\Delta$ can be related to the momentum of nucleon and pion from the decay of $\Delta$ in its rest frame. The larger the $m^*_\Delta$ is, the larger the $q$ is.

The form factors are adopted to effectively consider the contribution from high-order terms and the finite size of baryons \cite{Huber1994,Vetter1991}, which read
\begin{eqnarray}
\label{factorn}
&&F_N (t^*)=\frac{\Lambda_N^2}{\Lambda_N^2-t^*}  exp\left(-b\sqrt{s^*-4m_N^{* 2}}\right)\\
&&F_{\Delta}(t^*)=\frac{\Lambda_{\Delta}^2}{\Lambda_{\Delta}^2-t^*}.
\end{eqnarray}
Here $F_N (t^* )$ is the form factor for nucleon-meson-nucleon , and $F_\Delta (t^*)$  for  nucleon-meson-$\Delta$ coupling, $b$=0.046 GeV$^{-1}$ for both $\rho NN$ and $\pi NN$.
The cutoff parameter $\Lambda_{\pi N N}\approx 1$ GeV for all selected three parameter sets, i.e.,  NL$\rho\delta$, DDME$\delta$ and  DDRH$\rho\delta$. $\Lambda_{\rho N N}$ and $\Lambda_{\pi N \Delta}$ are determined by best fitting the data of $NN\rightarrow N\Delta$ cross section in free space \cite{Baldini1987} ranging from $\sqrt{s}$=2.0 to 5.0 GeV. In the Table.~\ref{table-para}, $\Lambda_{\rho N \Delta}$ is determined based on the relationship $\Lambda_{\rho N \Delta}=\Lambda_{\rho NN}\frac{\Lambda_{\pi N\Delta}}{\Lambda_{\pi NN}}$ as in \cite{Huber1994}.

\section{Results and discussions}
\label{xs}

\subsection{Cross section and its medium correction}

\begin{figure}[htbp]
\begin{center}
\includegraphics[scale=0.31]{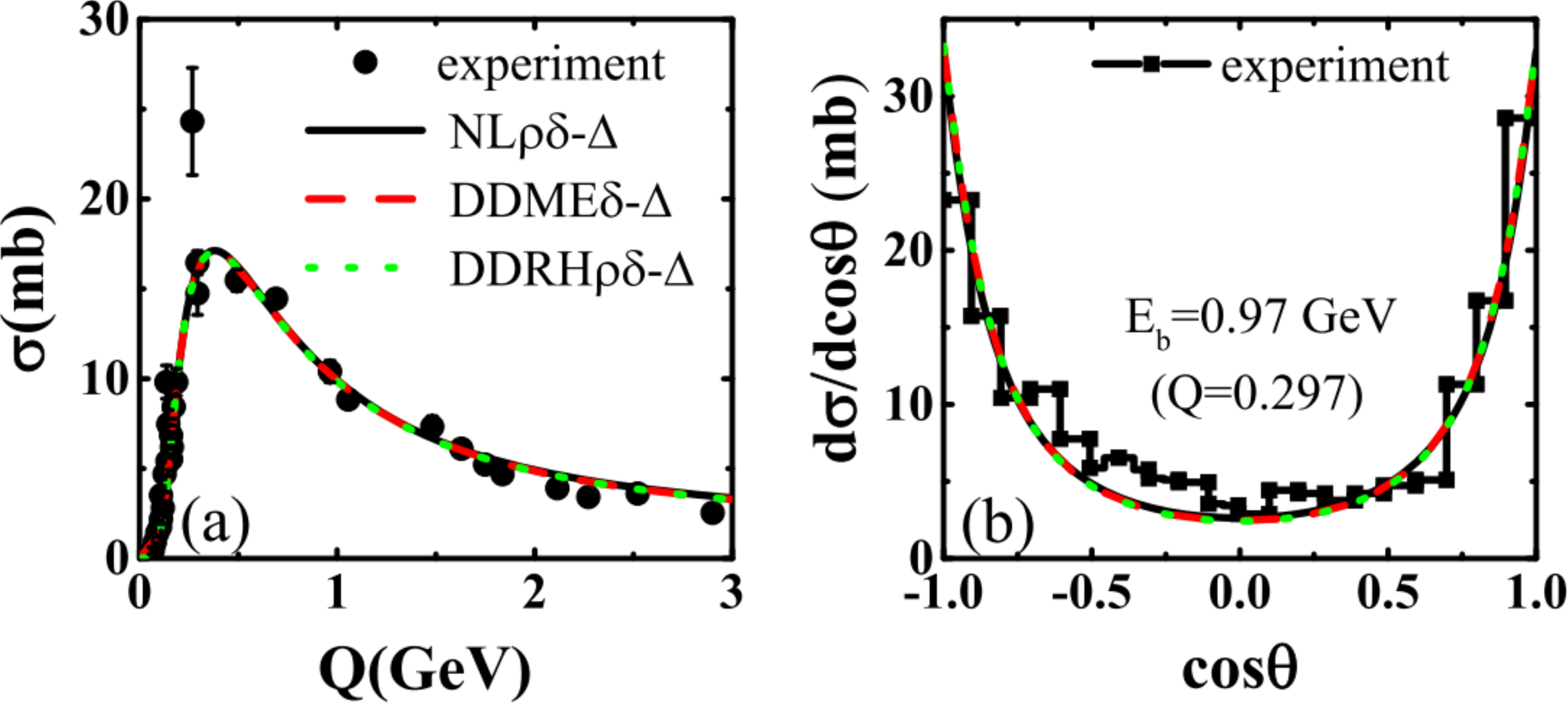}
\caption{(Color online) (a) $\sigma^*_{pp\rightarrow n\Delta^{++}}$ as a function of $Q$ for for   NL$\rho\delta$-$\Delta$, DDME$\delta$-$\Delta$ and DDRH$\rho\delta$-$\Delta$ in free space , the experimental data are from \cite{Baldini1987};(b) $\frac{d\sigma}{d cos\theta}$ as a function of $cos\theta$ at beam energy $E_{b}=0.97$ GeV, the experimental data from \cite{Bugg1964}. The lines with different colors correspond to different parameter sets. }\label{fig3}
\end{center}
\end{figure}

Figure~\ref{fig3} (a) shows the calculated $\sigma^*_{pp\rightarrow n\Delta^{++}}$ as a function of $Q$, and \ref{fig3} (b) shows $\frac{d\sigma^*}{d cos\theta}$ at the beam energy of 0.97 GeV  in free space, respectively.
$Q$ represents the kinetic energy above the pion production threshold energy $\sqrt{s_{\text{th}}}=m^*_{N_3}+m^*_{\Delta,\text{min}}+\Sigma^0_{N_3}+\Sigma^0_\Delta$, which is defined as
\begin{eqnarray}
\label{eq:qeff}
Q&=&\sqrt{s_{\text{in}}}-\sqrt{s_{\text{th}}}\\\nonumber
&=&E^*_{N_1}+E^*_{N_2}+\Sigma^0_{N_1}+\Sigma^0_{N_2}\\\nonumber
&&-m^*_{N_3}-m^*_{\Delta,\text{min}}-\Sigma^0_{N_3}-\Sigma^0_{\Delta}\\\nonumber
&\simeq &(E^*_{N_1}-m^*_{N_1})+(E^*_{N_2}-m^*_{N_2})\\\nonumber
&&+m_{N_1}+m_{N_2}-m_{N_3}-m_{\Delta,\text{min}}\\\nonumber
&&+\Delta\Sigma^S+\Delta\Sigma^0
\end{eqnarray}
here $\Delta\Sigma^S=\Sigma^S_{N_1}+\Sigma^S_{N_2}-\Sigma^S_{N_3}-\Sigma^S_{\Delta}$.
The black circles and squares correspond to experimental data \cite{Baldini1987,Bugg1964}.
The black solid line, dashed and dotted lines are the results for NL$\rho\delta$-$\Delta$, DDME$\delta$-$\Delta$ and DDRH$\rho\delta$-$\Delta$, respectively.
In order to investigate the impacts of different effective Lagrangian parameter sets on the in-medium $NN\rightarrow N\Delta$ cross section, all the selected parameter sets are adjusted to reproduce the experimental data of $NN\rightarrow N\Delta$ cross sections and their differential cross sections  at $E_b=$0.97 GeV where the data of differential cross section we can find.

\begin{figure}[htbp]
\begin{center}
    \includegraphics[scale=0.53]{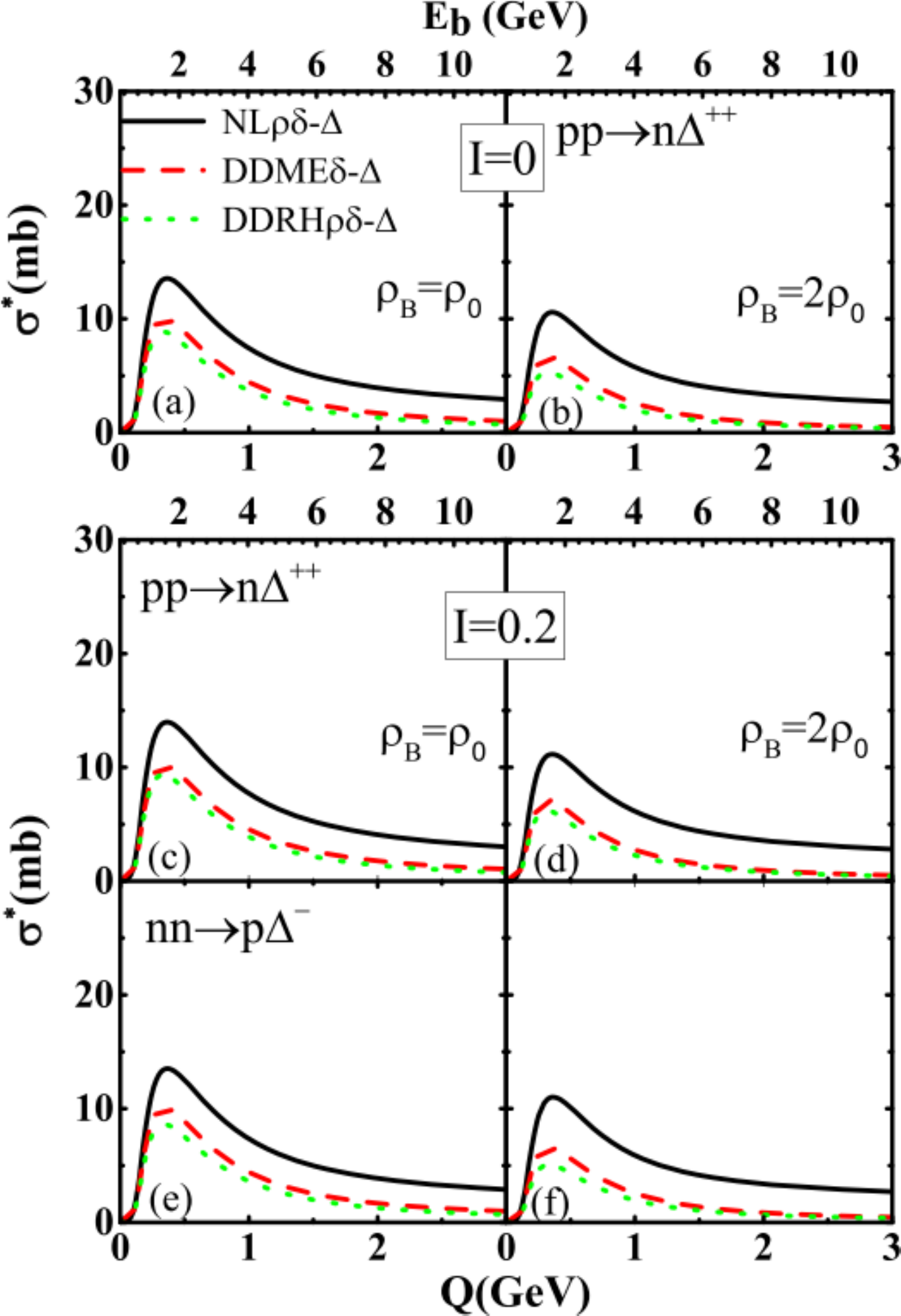}
    \caption{(Color online) $\sigma^*_{NN\rightarrow N\Delta}$ as a function of $Q$, (a) and (b)  for symmetric nuclear matter $I=0$;  (c)-(f) for asymmetric nuclear $I$=0.2.}\label{fig5}
\end{center}
\end{figure}

Fig.~\ref{fig5} (a) and (b) present the results of $\sigma^*_{pp\rightarrow n\Delta^{++}}$ at $\rho_0$ and $2\rho_0$ in symmetric nuclear matter for different parameter sets. The black solid line, red dashed and green dotted lines are the results for NL$\rho\delta$-$\Delta$, DDME$\delta$-$\Delta$ and DDRH$\rho\delta$-$\Delta$, respectively.
The values of $\sigma^*_{pp\rightarrow n\Delta^{++}}$  depend on the selected parameter sets. The NL$\rho\delta$-$\Delta$ predicts the largest in-medium $NN\rightarrow N\Delta$ cross section among three parameter sets, and  $\sigma^*_{\text{NL}\rho\delta\text{-}\Delta}>\sigma^*_{\text{DDME}\delta\text{-}\Delta}>\sigma^*_{\text{DDRH}\rho\delta\text{-}\Delta}$, especially at $2\rho_0$. The difference between $\sigma^*_{\text{DDME}\delta\text{-}\Delta}$ and $\sigma^*_{\text{DDRH}\rho\delta\text{-}\Delta}$ is comparatively small due to their slight difference between the effective masses as shown in Table~\ref{table-para}.
This can be understood from the formula of in-medium $NN\rightarrow N\Delta$ cross section, such as Eq.(\ref{eq:xsnd3}), where the values of cross section monotonically increase with the effective mass of nucleon and $\Delta$. The larger the effective mass is, the larger the cross section is.
Similar to the symmetric nuclear matter, the in-medium $NN\rightarrow N\Delta$ cross section in isospin asymmetric nuclear matter also has $\sigma^*_{\text{NL}\rho\delta\text{-}\Delta}>\sigma^*_{\text{DDME}\delta\text{-}\Delta}>\sigma^*_{\text{DDRH}\rho\delta\text{-}\Delta}$, which can be observed in Fig.~\ref{fig5} (c)-(f), where the $\sigma^*_{pp\to n\Delta^{++}}$ and $\sigma^*_{nn\to p\Delta^{-}}$ at $\rho_0$ (left panels) and  $2\rho_0$ (right panels) for isospin asymmetry $I$=0.2 are shown as an example.

Based on our discussion in \cite{Cui2018}, the in-medium $NN\rightarrow N\Delta$ cross section is split in isospin asymmetric nuclear matter due to the effective mass splitting for nucleons and $\Delta$s. The values of in-medium cross sections of $pp\rightarrow n\Delta^{++}$, $pp\rightarrow p\Delta^{+}$, $pn\rightarrow n\Delta^{+}$, $pn\rightarrow p\Delta^{0}$, $nn\rightarrow n\Delta^{0}$, and $nn\rightarrow p\Delta^{-}$  do not satisfy the Clebsch-Gordan coefficients as the free space scenario. It can be understood from the expression of matrix element in Eq.~\ref{eq:xsnd3}. For example, if there is no isospin splitting for nucleon and $\Delta$ effective mass,  the difference of $|\mathcal{M}|^2$ between the different channels come from $I_d^2$ or $I_e^2$ because the terms contains the $m_N^*$, $m^*_\Delta$, $t^*$ in $|\mathcal{M}|^2$ have the same contributions to different channels. But in the isospin asymmetric nuclear matter, there is isospin splitting on the nucleon and $\Delta$ effective mass, and it causes different values of $m_N^*$, $m^*_\Delta$, $t^*$ in $|\mathcal{M}|^2$ in addition to $I_d^2$ and $I_e^2$ for different channels.

\begin{figure}[htbp]
\begin{center}
\includegraphics[scale=0.33]{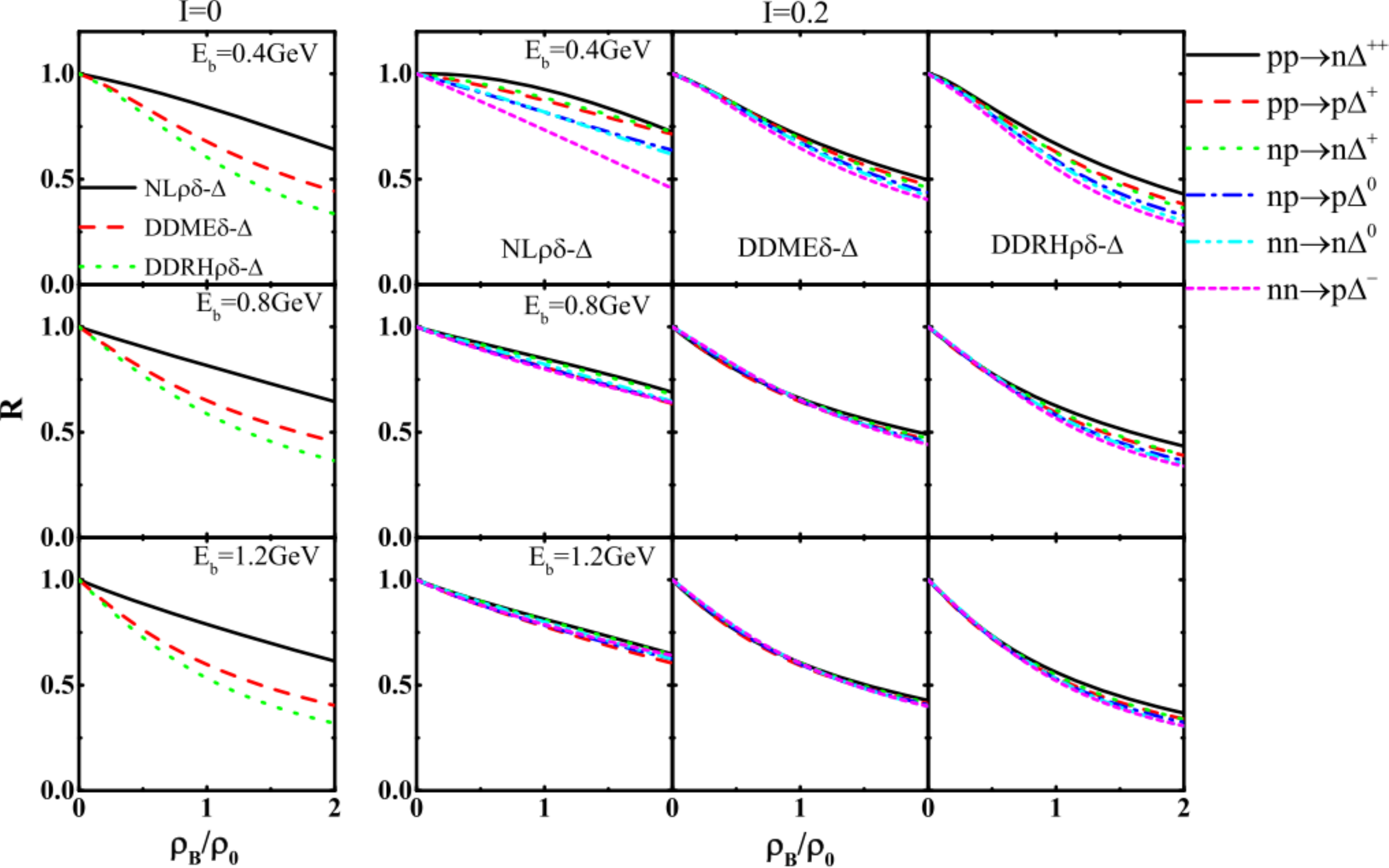}
\caption{(Color online) The medium correction factor $R=\sigma^{*}/\sigma^{\text{free}}$ of different channels (with different color) as the function of density for $E_b$=0.4, 0.8 and 1.2 GeV  ($Q$= 0.052, 0.227 and 0.389 GeV) for different parameter sets. Left three panels are for symmetric nuclear matter ($I$=0), right nine panels for asymmetric nuclear matter ($I$=0.2) .}\label{fig6}
\end{center}
\end{figure}

In the left panels of Fig.~\ref{fig6}, we present the $R$ ratios in the symmetric nuclear matter. The upper, middle, and bottom panels correspond to the results for different beam energies or $E_b$=0.4 ($Q$=0.052 GeV), 0.8 ($Q$=0.227 GeV), and 1.2 GeV ($Q$=0.389 GeV), respectively. The different channels have the same in-medium correction factor $R$, and their values are decreased with the increasing of the density. It is consistent with the work from \cite{Larionov2003, QingfengLi2017}. Similar to the dependence of cross section on the parameter sets, $R_{\text{NL}\rho\delta\text{-}\Delta}>R_{\text{DDME}\delta\text{-}\Delta}>R_{\text{DDRH}\rho\delta\text{-}\Delta}$.

For isospin asymmetric nuclear medium, in the right panels of Fig.~\ref{fig6}, the $R$ ratios obtained with the selected parameter sets also decrease as the function of density, and they are split according to the different isospin state of collision channels. Near the threshold energy, the $R$ values clearly depend on the channel of $NN\rightarrow N\Delta$ and $R(pp\rightarrow n\Delta^{++})>R(Np\rightarrow N\Delta^{+})>R(Nn\rightarrow N\Delta^{0})>R(nn\rightarrow p\Delta^{-})$, here $N=n$ or $p$. The amplitude of the splitting mainly attributes to the effective mass splitting of nucleon and $\Delta$, which are presented in Table~\ref{table-para} via the effective mass changes between  incoming and outgoing particles, i.e., $\Delta\Sigma^S$,  and the effective energy changes, i.e., $\Delta\Sigma^0$ for different channels. In the calculation of the in-medium $NN\to N\Delta$ cross section, the values of $\Delta \Sigma^S$  and $\Delta \Sigma^0$ provide the opposite contribution on their isospin effects through $Q$.
Near the threshold ($E_b\approx $0.4 GeV), the  $R$ values are mainly effected by the effective mass changes $\Delta \Sigma^S$ and effective energy changes $\Delta \Sigma^0$. With the beam energy increasing up to 0.8 GeV, the splitting of $R$ among the different channels of $NN\rightarrow N\Delta$ tends to vanish because the contributions from scalar and vector self energies become relatively smaller than the kinetic energy part.

\begin{figure}[htbp]
\begin{center}
    \includegraphics[scale=0.24]{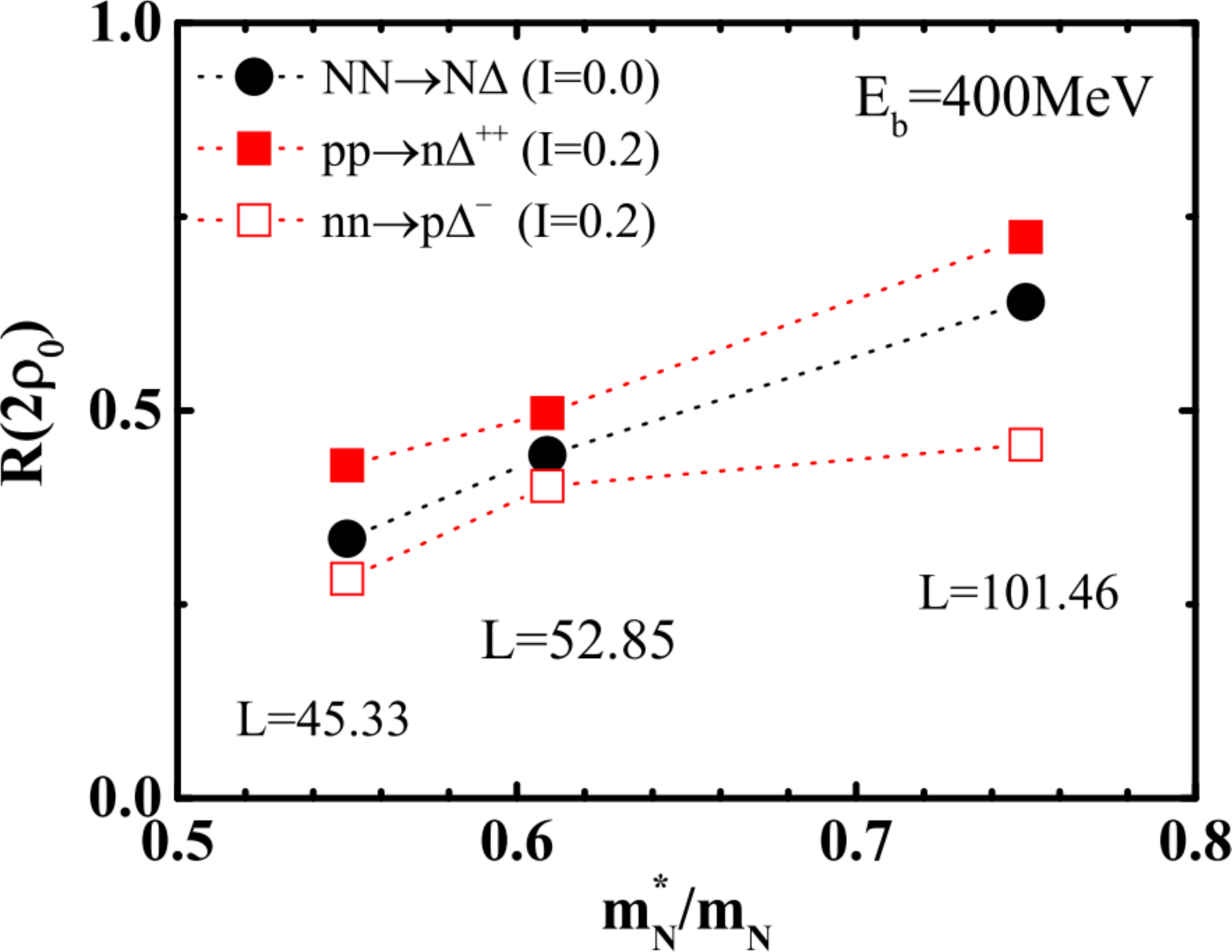}
    \caption{(Color online) The medium correction factor R at $\rho_B=2\rho_0$ in $E_b$=0.4 GeV for different parameter sets, i.e., NL$\rho\delta$-$\Delta$, DDRH$\rho\delta$-$\Delta$ and  DDME$\delta$-$\Delta$, the unit of $L$ is MeV. }\label{figRL}
\end{center}
\end{figure}

Near the threshold energy, the splitting of $R$ is larger in NL$\rho\delta$-$\Delta$ than that in DDRH$\rho\delta$-$\Delta$ and  DDME$\delta$-$\Delta$ due to the stronger nucleons and $\Delta$s effective mass splitting in NL$\rho\delta$-$\Delta$. The splitting of $R$ for different channels  vanish at $E_b>$ 0.8 GeV for  all  parameter sets, but the reduction of in-medium correction follows $R_{\text{NL}\rho\delta\text{-}\Delta}>R_{\text{DDME}\delta\text{-}\Delta}>R_{\text{DDRH}\rho\delta\text{-}\Delta}$  that is related to the decreasing of effective masses for the three parameter sets in Table~\ref{table-para}.   It can be seen more clearly in Fig.~\ref{figRL},  in which the $R(2\rho_0)$ increases with $m^*_{N}/m_{N}$ (or $L$) increasing.
It hints that adjusting the medium correction factor $R$ in transport models should also consider the stiffness of isospin asymmetric nuclear equation of state simultaneously. However, the concrete relationship between the medium correction factor and stiffness of symmetry energy still needs lots of work, for example, by analyzing the proposed hundreds of RMF parameters.

\begin{figure}[htbp]
\begin{center}
\includegraphics[scale=0.36]{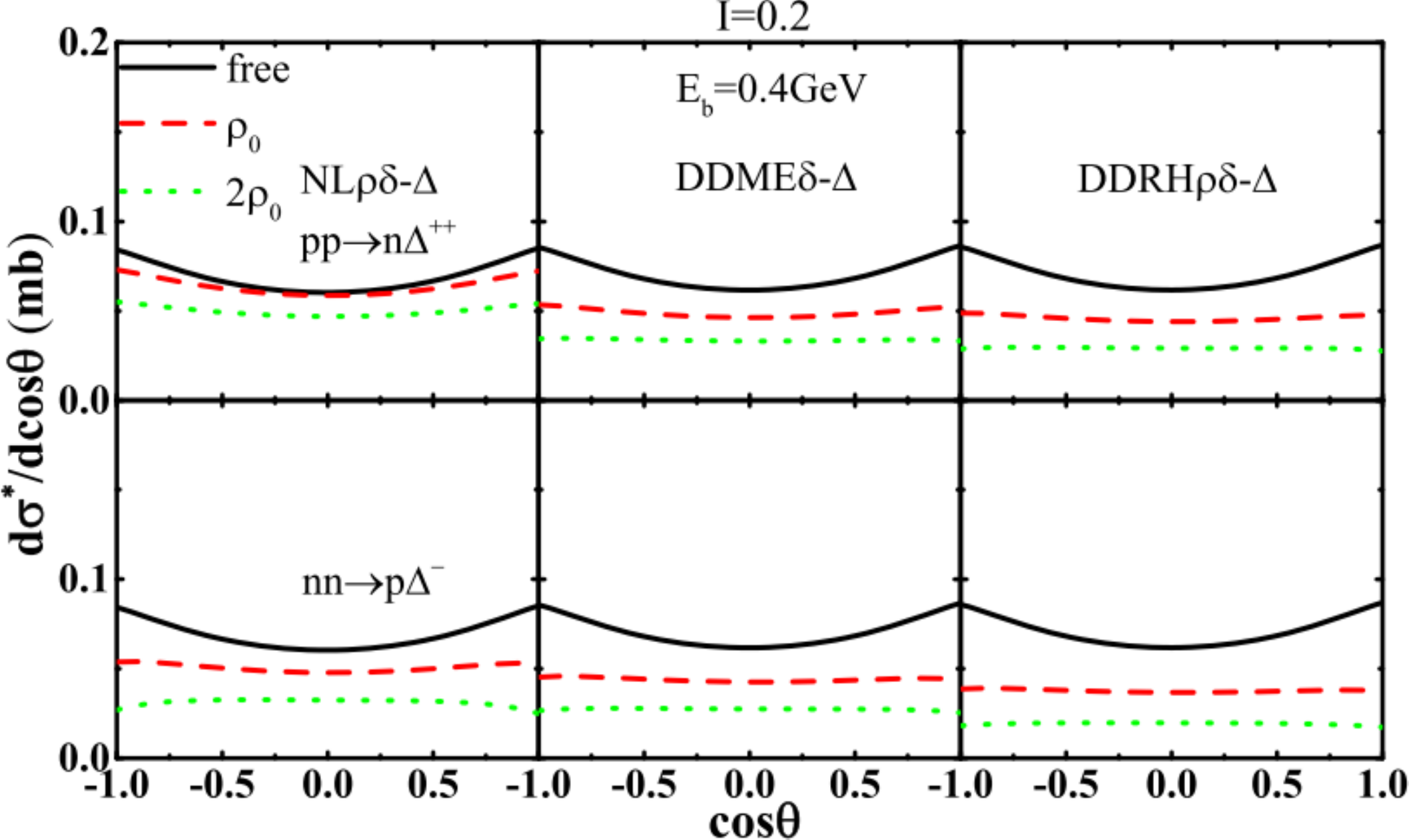}
    \caption{(Color online)  $d\sigma^*/d cos\theta$ for $pp\to n \Delta^{++}$ and $nn\to p \Delta^{-}$ channels as a function of $cos\theta$ at the beam energy of 0.4 GeV. The lines with different colors correspond to $\rho_B=0, \rho_0, 2\rho_0$ in asymmetric nuclear matter ($I$=0.2). The panels from left to right refer to the results obtained with NL$\rho\delta$-$\Delta$, DDME$\delta$-$\Delta$, and DDHR$\rho\delta$-$\Delta$.}\label{fig7}
\end{center}
\end{figure}

Since the differential $NN\to N\Delta$ cross section determines the scattering angle for colliding particles in transport models, the discussion of the medium effects on the differential cross sections for $NN\to N\Delta$ is also an indispensable part.
A parameterized form of differential cross sections from experimental data\cite{Cug96} is usually used in many codes without considering the medium correction effects. Recently, Wang et.al. \cite{Wang2016} have tried to understand the influence of the different forms of differential cross section on the elliptical flow in the ultrarelativlstic  quantum molecular
dynamics model (UrQMD)
simulations, and their results show it could influence the nuclear stopping power, direct and elliptic flow at high beam energies. It also stimulated the theoretical understanding of the in-medium differential $NN\to N\Delta$ cross sections which are needed for developing the isospin dependent transport codes.
The in-medium differential cross sections become more isotropic with increasing density for the elastic $NN$ collisions \cite{Zhanghongfei2010}, and the similar behaviors have been found in the $NN\to N\Delta$ differential cross section in symmetric nuclear matter\cite{Larionov2003}. 
Our calculations also confirm the conclusion that the differential cross section for $NN\to N\Delta$ tends to be more isotropic for all the parameter sets we used in the case of symmetric nuclear medium, especially at the twice normal density near the threshold energy. 
Furthermore, the same behaviour of the in-medium $NN\to N\Delta$ differential cross sections can be found in asymmetric matter. As shown in Fig.~\ref{fig7}, we present the results of $pp\to n \Delta^{++}$ and $nn\to p \Delta^{-}$ channels at $E_b=0.4 $ GeV as an example. The medium correction of the differential cross section is strong, and it mainly appears at forward and backward region, i.e.,  $\theta_{c.m.}<60^\circ$ and $\theta_{c.m.}>120^\circ$.
When the beam energy is higher, the medium correction effects become weaker around the $\theta_{c.m.}=90^\circ$, but it still exists at forward and backward region.

\section{Summary}
\label{summary}
In summary, we have studied the in-medium $NN\rightarrow N\Delta$ integrate and differential cross sections in isospin asymmetric nuclear medium within the one-boson exchange model. Three different interaction parameter sets, which involves $\rho$ and $\delta$ mesons, are adopted in this work. Our calculations show that $\sigma^*_{NN\rightarrow N\Delta}$ decreases with the density increasing, and the in-medium differential cross sections become more isotropic with density increasing near the threshold energy for all the selected parameter sets. At the given density, the medium correction factor $R$ decreases with the effective mass decreasing, or with decreasing of the slope of symmetry energy. This information is useful for mimic the deficiency of transport models, where the mean field and in-medium nucleon-nucleon cross section are adjusted separately in order to fit the data. By considering the relationship between the in-medium $NN\to N\Delta$ cross sections and slope of symmetry energy in the transport model calculations,  it could reduce the ambiguity of the constrains on either EOS or in-medium $NN\to N\Delta$ cross section through the comparison with heavy ion collisions data.

To concrete the relationship between the EOS and in-medium $NN\to N\Delta$ cross section, further analysis on the proposed RMF parameter sets are required. For example, there are 263 RMF parameter sets\cite{Dutra14} and most of them only include $\sigma$, $\omega$ and $\rho$ mesons. In the parameter sets with $\sigma$, $\omega$ and $\rho$ mesons, the relation we found in above could be modified because the isospin splitting of $R$ is only caused by isospin splitting of effective energy by $\rho$ meson. The work in this direction will be interesting and helpful for reliable extracting the EOS or in-medium NN cross section through transport models in the future.

\acknowledgments
This work has been supported by National Natural Science Foundation of China under Grants No. 11875323, No. 11875125, No. 11475262, No. 11365004, No. 11375062, No. 11790323,11790324, and No. 11790325 and the National Key R\&D Program of China under Grant No. 2018 YFA0404404.

\end{document}